\documentclass[useAMS,usenatbib]{mn2e}

\title[Loop orbits]
{Orbits in elementary, power-law galaxy bars: 1. Occurence and role of single loops}
\author[C. Struck] 
{Curtis Struck  \thanks{E-mail: curt@iastate.edu} \\
Department of Physics and Astronomy, Iowa State University, Ames, IA, 50014 USA}
\usepackage{amssymb}
\usepackage{amsmath}
\usepackage{graphicx}
\usepackage{multirow} 
\def\aap{{ A\&A}}

\def\aj{{AJ}}

\def\apj{{ApJ}}

\def\mnras{{MNRAS}}

\begin{document}
\date{\today}

\pagerange{\pageref{firstpage}--\pageref{lastpage}} \pubyear{0000}

\maketitle

\label{firstpage}
\begin{abstract}

Orbits in galaxy bars are generally complex, but simple closed loop orbits play an important role in our conceptual understanding of bars. Such orbits are found in some well-studied potentials, provide a simple model of the bar in themselves, and may generate complex orbit families. The precessing, power ellipse (p-ellipse) orbit approximation provides accurate analytic orbit fits in symmetric galaxy potentials. It remains useful for finding and fitting simple loop orbits in the frame of a rotating bar with bar-like and symmetric power-law potentials. Second order perturbation theory yields two or fewer simple loop solutions in these potentials. Numerical integrations in the parameter space neighborhood of perturbation solutions reveal zero or one actual loops in a range of such potentials with rising rotation curves. These loops are embedded in a small parameter region of similar, but librating orbits, which have a subharmonic frequency superimposed on the basic loop. These loops and their librating companions support annular bars. Solid bars can be produced in more complex potentials, as shown by an example with power-law indices varying with radius. The power-law potentials can be viewed as the elementary constituents of more complex potentials. Numerical integrations also reveal interesting classes of orbits with multiple loops. 

In two-dimensional, self-gravitating bars, with power-law potentials, single loop orbits are very rare. This result suggests that gas bars or oval distortions are unlikely to be long-lived, and that complex orbits or three-dimensional structure must support self-gravitating stellar bars.

\end{abstract}


\section{Introduction}

The idea that galaxy bars are built on a skeleton of simple closed orbits, elongated along the bar (i.e., the $x_1$ family), and fleshed out by similar orbits with constrained librations \citep{ly79}, is conceptually simple, and popular. \citet{at13} states - ``The bar can then be considered as a superposition of such orbits, ..., which will thus be the backbone of the bar." Images of nested sets of such orbits derived from numerical models reinforce that idea (see \citealt{at92}, \citealt{bt08}). Nonetheless, there are very few analytic models of stellar bars and oval distortions in galaxies, composed of simple, nested orbits (see \citealt{co02}, \citealt{bt08}). This is unfortunate since such models could facilitate the study of bars, and advance our understanding of them. 

Furthermore, much of the study of orbits in numerical simulations has focussed on the special cases of fixed potentials with bars of Ferrers (e.g., \citealt{at92}) or Freeman type \citep{fr66a, fr66b, fr66c}. \citet{wi17} point out that, because of their homogeneous density profiles, these models are not especially realistic. These latter authors study a family of very different models where the bar components are represented by thin, dense needles. Indeed, they opine that``models of bars ... remain rather primitive'' and ``there is ample scope for the development of new models...''

In their models of weak bars \citet{wi17} do find that, as in the classic picture, simple, loop (type $x_1, x_4$) dominate. However, their models of strong bars, are dominated by much more complex `propeller' orbits. Families of complex and chaotic orbits are found in many numerical simulations with either fixed or self-consistent potentials (e.g., \citealt{se93}, \citealt{er14}, \citealt{ma14}, \citealt{ju15}, \citealt{va15}, \citealt{ga16}), and those families are likely to be just as important a constituent of bars as simple loop orbits. Thus, the question arises of whether the classic picture of bars as nested loop orbits has any great relevance beyond special cases or illustrative toy models? 

On the other hand, gas clouds in bars cannot pursue complex orbits without generating shocks and strong dissipation. Gas may be quickly expelled from strong bars dominated by complex orbits, but may play an important role in weak or forming bars. Thus, beyond generalizing the \citet{wi17} models, it would be useful to know when, and in what potentials simple, nested, loop orbits can dominate the bar.

Galaxies, or even limited radial regions in galaxy discs, have a wide range of potential forms. It would be useful to find relationships between the structure of potentials (symmetric and asymmetric) and the orbit types, especially simple orbit types, that they support. This is another area where our knowledge ``remain(s) rather primitive''. In this paper I will undertake a modest exploration of this territory by studying the simplest orbits in simple power-law potentials. More realistic potentials may be decomposed into sums of power-law potential approximations, and we may expect that individual terms in these sums will bring their corresponding orbits into regions where they dominate. 

There are several ways to study closed orbits in bars (see \citealt{be00}, \citealt{bt08}. Perhaps, the most direct method, is to seek them in numerical models (e.g, \citealt{co89}, \citealt{at92}, \citealt{mi98}). A second method leverages action-angle variables in a perturbation formalism, which in limiting cases fits well with the epicyclic orbit approximation (e.g., \citealt{ly79}, \citealt{se93}, \citealt{bt08}, \citealt{se14}). In this paper we will use a related method, analytic (p-ellipse) orbit approximations in a perturbation approach. 

In \citet{st06} it was shown that a precessing power-law ellipse (p-ellipse) approximation is quite accurate up to moderate eccentricities in a wide range of power-law potentials. There are other good approximations available, e.g., the Lambert W function discussed in \citep{va12}, but p-ellipses are especially simple. In a later work \citep{st15a} it was found, that with simple modifications, i.e., to the precession frequencies, p-ellipse approximations can also approximate high eccentricity orbits remarkably well. Because of this frequency modification there is a continuum of Lindblad resonances parametrized by eccentricity for highly eccentric orbits. Ensembles of eccentric resonant orbits of different sizes, excited impulsively, could have equal precession periods and make up the backbone of kinematic bars or spiral arms with constant pattern speeds in symmetric halo potentials \citep{st15b}. This idea motivated the work below, but we will see that in many power-law potentials with a bar component, nearly radial, single loop orbits in the bar frame either do not exist or are very small.)\ 

The \citet{st15b} paper did not address the question of whether these or or other simple closed orbits also exist in potentials with a fixed non-axisymmetric component, e.g., due to a prolonged tidal component or an oval or bar-like halo. To use approximate p-ellipse orbits to investigate this it must first be shown when, or under what conditions, p-ellipses can approximate orbits in non-axisymmetric gravitational potentials. It will be demonstrated below that in the case of simple, closed loop orbits the answer is the same as in the case of symmetric potentials - the p-ellipse approximation is again quite accurate up to moderate eccentricities in a wide range of potentials (Sec. 4).

It will also be shown by example, that in the immediate neighborhood of resonant loop orbits, there exist other orbits that are very similar, but modestly librating (Sec. 4). On average, these orbits can also be described by p-ellipses, and ultimately may be more completely approximated by p-ellipse with added frequencies to represent the libration (see Sec. 3.2). The parameter space near the simple resonant loop is evidently densely populated with closed versions of these librating orbits, and they might be used to form the backbone of a model bar (Sec. 5). This suggestion is much as proposed by \citet{ly79}, also see \citet{co77}, and \citet{ly96} for discussions of resonant orbits and bar formation. (\citet{ly79}, in an appendix, also describes a wider range of orbits that would fit into his formalism.)

We will see in Sec. 4 and 5 that many potentials with symmetric and barred components represented by single power-laws do not have more than one closed ($m=2$) loop orbit. Even when accompanied by their librating family of nearby orbits, we would only expect hollow, annular stellar bars to exist in these cases. A potential consisting of multiple power-law parts, each dominating in successive annular ranges, can produce a nested series of closed orbits, and their librating companions. This can make a more robust bar, see Sec. 5 and Fig. 8.

The excitation of resonant orbits by tidal disturbances or asymmetric halos may generate self-gravitating bars or waves (\citealt{no87, no88, ba91}). It is not clear, however, that as the bars grow to nonlinearity, and acquire significant self-gravity, whether the simple, closed orbits will continue to exist, or if they can be arranged to form a stable, self-gravitating bar. I.e., whether the Poisson equations, as well as the equations of motion, can be approximately solved by ensembles of simple, loop orbits and modestly librating orbits. 

In fact, we will see in Sec. 6 that the simple planar, p-ellipse approximation at second order has very few solutions with the additional Poisson constraints. As discussed in the final two sections, these results imply that long-lived, self-consistent bars or oval distortions cannot have a substantial gas component, because there would be strong dissipation. When such bars are made of stars, essentially all orbits must librate, or have complex multi-loop forms, as seen in published simulations.

\section{Basic equations and p-ellipse approximations}
\subsection{Basic equations}

In this work we only consider orbits in the two-dimensional central plane of a galaxy disc, and generally adopt a symmetric, power-law, halo potential of the form, 

\begin{equation}
\label{eqa}
\Phi = \frac{-GM_{\epsilon}}{2{\delta}\epsilon}  
\left(\frac{\epsilon}{r} \right) ^{2\delta}.
\end{equation}

\noindent In addition we will include a non-axisymmetric (bar) part of the potential of the simple form,

\begin{equation}
\label{eqb}
\Phi_b = \frac{-GM_{\epsilon}}{\epsilon}  
\left(\frac{\epsilon}{r} \right) ^{2\delta_b}
{e_b} cos \left( 2(\phi - \phi_o) + \Omega_b t \right) ,
\end{equation}

\noindent where $r$ and $\phi$ are the radial and azimuthal coordinates in the disc, $e_b$ is an amplitude parameter of the asymmetric potential, $\delta$ and $\delta_b$ give the radial dependence of the symmetric and non-axisymmetric potentials, and $\Omega_b$ is the rotation frequency of the latter. The scale length is $\epsilon$ and $M_\epsilon$ is the halo mass contained between the radius $r = \epsilon$ and some minimum radius. The above is a very simple form for a bar potential, with relatively few parameters, and no characteristic length (e.g., cutoff radius). 

Then the equations of motion for stars orbiting in the disc with the adopted potentials are, 

\begin{equation*} \begin{split}
\ddot{r} = \frac{-GM_{\epsilon}}{\epsilon^2}
\left( \frac{\epsilon}{r} \right) ^{1+2\delta} - 
2{\delta_b}e_b \frac{GM_{\epsilon}}{\epsilon^2}
\left( \frac{\epsilon}{r} \right) ^{1+2\delta_b}\\
\times \cos{\left( 2(\phi - \phi_o) + \Omega_b t \right)} 
+ r \dot{\phi}^2, 
\end{split} \end{equation*}
\begin{equation} \begin{split}
\label{eqc}
\ddot{\phi} + \frac{2\dot{r}\dot{\phi}}{r} =
-\frac{1}{r^2} \frac{\partial \Phi}{\partial \phi} =
-2e_b \frac{GM_{\epsilon}\Omega_b}{\epsilon^3}
\left( \frac{\epsilon}{r} \right)^{2+2\delta_b}\\ 
\times \sin{\left( 2(\phi - \phi_o) + \Omega_b t \right)}.
\end{split} \end{equation}

Next, we derive dimensionless forms of these equations by substituting the dimensionless (overbar) variables and dimensionless constants defined as, 

\begin{equation}
\label{eqd}
\bar{r} = r/\epsilon,\ \bar{t} = t/\tau,\ 
c = \frac{GM_\epsilon \tau^2}{ \epsilon^3},\ 
c_b = c e_b.
\end{equation}

\noindent For additional simplification we will set the value of the timescale to $\tau^{-2} = \frac{GM_\epsilon}{\epsilon^3}$, so that $c = 1.0$. Despite this choice, we will carry the $c$ factor through much of the analysis below for clarity. Then the dimensionless equations of motion are, 

\begin{multline*}
\ddot{\bar{r}} = -c 
\bar{r}^{-\left( {1+2\delta}\right)} - 
2{\delta_b}c_b
\bar{r}^{-\left( {1+2\delta_b}  \right)}
\cos{\left( 2(\bar{\phi} - \phi_o) + \Omega_b t \right)} + 
\bar{r} \dot{\bar{\phi}}^2, 
\end{multline*}
\begin{equation}
\label{eqe}
\ddot{\bar{\phi}} + \frac{2\dot{\bar{r}}\dot{\bar{\phi}}}{\bar{r}} =
-2c_b \Omega_b
\bar{r}^{-\left(  {2+2\delta_b} \right)} 
\sin{(\left( 2(\bar{\phi} - \phi_o) + \Omega_b t \right)}.
\end{equation}

\noindent \textit{Henceforth we will omit the overbars and assume all variables are dimensionless.} We will also assume that the initial value of the azimuth ($\phi_o$) is zero. 

The next step towards a more workable set of equations is to go into a reference frame rotating with the bar or pattern speed, $\Omega_b$. In this frame the dimensionless radii are the same, and in terms of the previous values, the azimuthal coordinates are $\phi' = \phi - \Omega_b t$. We will henceforth drop the primed notation, so that the equations of motion in the rotating frame are, 

\begin{multline*}
\ddot{r} = -c 
r^{-\left( {1+2\delta}\right)} - 
2{\delta_b}c_b
r^{-\left( {1+2\delta_b}  \right)}
\cos{(2\phi)} + 
r \left( \dot{\phi} + \Omega_b \right)^2, 
\end{multline*}
\begin{equation}
\label{eqee}
\ddot{\phi} + \frac{2\dot{r} \left( \dot{\phi} +  \Omega_b \right)}{r}
 = -2c_b 
r^{-\left(  {2+2\delta_b} \right)} 
\sin(2\phi) 
\end{equation}

\noindent (see e.g., \citet{bt08}, equations 3.135a,b).

\subsection{p-ellipse approximations}

As described in the Introduction, we seek approximate solutions of these equations, of the form,

\begin{equation}
\label{eqf}
\frac{1}{r} = \frac{1}{p} \left[ 1 +
e \cos \left( m{\phi} \right) 
\right]^{\frac{1}{2} + \delta},
\end{equation}

\noindent which were studied in \citet[Paper 1]{st06}, named precessing, power-law ellipses, or `p-ellipses', and found to be quite accurate despite their simplicity (for other approximations see \citealt{va12}). Here the orbital scale is given by the semi-latus rectum $p$, $m$ is a frequency ratio, and $e$ is the eccentricity parameter. Note that while the form of equation \eqref{eqf} is that same as in \citet{st06}, and subsequent p-ellipse papers, the physical meaning of the $m$ parameter is different in the rotating coordinate system, though still a function of the ratio of precession and orbital frequencies. In the following we will focus on the case where this solution is in resonance with the bar driving force, i.e., with $m = 2$.

If such solutions can provide accurate approximations, as in the case of symmetric potentials, then they demonstrate continuity with orbits of the purely symmetric part of the potential (since parameters from the bar potential are not included). They might also provide a useful tool for studying orbit transformation in the process of bar formation. However, it is not \textit{a priori} clear how well the p-ellipse approximation will work for orbits that change their angular momenta over orbital segments (conserving it only over the whole period in the case of closed resonant orbits). 

Generally, equation \eqref{eqf} only yields closed or open-precessing loop forms except at high eccentricity. As detailed in \citet{st15a}, in the case of nearly radial orbits, the addition of a harmonic term (in $cos(2m\phi)$) to equation \eqref{eqf} significantly improves the accuracy of the orbit approximation. We will not pursue this refinement in the present paper, and to keep the algebra manageable, will generally neglect harmonic terms in the perturbation analyses below. However, it has been clear since the early work of \citet{ly79} that classes of orbits in bars can be described as liberating ovals. Analytic approximation of these forms requires more than a single frequency. Thus, we will explore the equations with a second frequency term ($m$) to get an approximate solution of the form, 

\begin{equation}
\label{eqfa}
\frac{1}{r} = \frac{1}{p} \left[ 1 +
e \cos \left( m{\phi} \right)  + c_2 e \cos \left( 2{\phi} \right)
+ c_x e^2 \cos \left( (2-m){\phi} \right) 
\right]^{\frac{1}{2} + \delta},
\end{equation}.

\noindent The new frequency is $m$ (here redefined and $\neq 2$), and the final term in square brackets must be included since such factors will be generated by cross terms in the equations of motion, so the solution must contain terms to balance them. The value of the frequency $m$ may be close to $2$. In such cases, the frequency $2-m$ will generally have a small value, and can approximate a subharmonic of the driving frequency. This can generate liberating, near resonant loop approximations to numerical orbits, as well as more complex forms. 

\section{Perturbation analyses}

In this section we develop the p-ellipse approximations to the orbits satisfying equations \eqref{eqee}, and derive the corresponding relations between the orbital parameters. Assuming that the orbits are not radial, so the eccentricity $e$ is a relatively small parameter, we can expand in that parameter. While the radial equation above has zeroth order terms, all the terms in the azimuthal equation are of first order and higher.  For reasonable accuracy up to moderate eccentricities, we carry out the expansion to second order. We consider the two approximate solutions given by equations \eqref{eqf} and \eqref{eqfa} separately in the following two subsections. Readers not interested in the details of these calculations may proceed to Sec. 4.

\subsection{Single frequency case}

 In this subsection we consider the perturbation expansion of the resonant solution equation \eqref{eqf}. The second order approximations to that p-ellipse solution and its first two time derivatives are,

\begin{equation} \begin{split}
\label{eqg}
\frac{r}{p} \simeq\  &1 - \left(\frac{1}{2}+\delta \right) e\ cos(2\phi) \\
&+ \frac{1}{2} \left(\frac{1}{2}+\delta \right) 
\left(\frac{3}{2}+\delta \right) e^2 cos^2(2\phi), 
\end{split} \end{equation}

\begin{equation} \begin{split}
\label{eqh}
\frac{1}{p} \frac{dr}{dt} &=
\frac{\dot{r}}{p} \simeq  \\
&\left[ \left(\frac{1}{2} + \delta \right) 2e\ sin(2\phi)
\left[ 1 - \left(\frac{3}{2}+\delta \right) e\ cos(2\phi) \right] \right]
\dot{\phi},
\end{split} \end{equation}

\begin{equation} \begin{split}
\label{eqi}
\frac{\ddot{r}}{p} \simeq &\left\{ \left(\frac{1}{2} + \delta \right) 
\left(\frac{3}{2} + \delta \right) 4 e^2 +
\left(\frac{1}{2} + \delta \right) 4 e cos(2\phi) \right. \\
&\left. - \left(\frac{1}{2} + \delta \right) \left(3 + \delta \right) 4 e^2
cos^2(2\phi) 
\right\} \dot{\phi}^2 \\
& + \left(\frac{1}{2} + \delta \right) 2 e\ sin(2\phi) \ddot{\phi}.
\end{split} \end{equation}

\noindent In the last of these equations an additional term in $e^2 \ddot{\phi}$ was dropped on the assumption that  $\ddot{\phi}$ is itself of first or higher order. This assumption will be confirmed below. This equation still contains one term in $\ddot{\phi}$. Substituting the expressions above for $r$ and $\dot{r}$ into the second of equations \eqref{eqee}, we obtain a first order approximation for $\ddot{\phi}$,

\begin{equation}
\label{eqj}
\ddot{\phi} \simeq -4 \left(\frac{1}{2} + \delta \right) 
\left( \dot{\phi} + \Omega_b \right) \dot{\phi}
e\ sin(2\phi) 
- \frac{2c_b}{p^{2(1+\delta_b)}} sin(2\phi),
\end{equation}

\noindent where we assume that $c_b$ is comparable to or less than $e$. This equation can then be substituted into equation \eqref{eqi}, and the resulting form substituted for $\ddot{r}$ in the first of equations \eqref{eqee}. The following approximations of the power-law terms can also be substituted.

\begin{equation}  \begin{split}
\label{eqk}
\left( \frac{p}{r} \right)^{1+2\delta}& \simeq 1 + 
2 \left(\frac{1}{2} + \delta \right)^2 e\ cos(2\phi) \\
& + \left(\frac{1}{2} + \delta \right)^2
\left[ 2\left(\frac{1}{2} + \delta \right)^2 - 1 \right]
\ e^2\ cos^2(2\phi),
\end{split} \end{equation}

\begin{equation}  \begin{split}
\label{eql}
&\left( \frac{p}{r} \right)^{2(1+\delta_b)} \simeq1 + 
2 \left(\frac{1}{2} + \delta \right) \left(\frac{1}{2} + \delta_b \right) e\ cos(2\phi) \\
& + \left(\frac{1}{2} + \delta \right) \left(\frac{1}{2} + \delta_b \right)
\left[ 2\left(\frac{1}{2} + \delta \right) \left(\frac{1}{2} + \delta_b \right) - 1 \right]\\
& \times\ e^2\ cos^2(2\phi),
\end{split} \end{equation}

After substituting equations \eqref{eqg} - \eqref{eql}, the radial equation in equation \eqref{eqee} yields a second order expression for $\dot{\phi}^2$, which is equivalent to that obtained from angular momentum conservation in symmetric potentials. We can generally approximate this variable, like the radius, in powers of $e\ cos(2\phi)$,

\begin{equation}
\label{eqm}
\dot{\phi} \simeq f_o + f_1 e cos(2\phi) + f_2 e^2 cos^2 (2\phi),
\end{equation}

\noindent where the $f_i$ are constant coefficients. With this final substitution, the radial equation yields a constraint equation at each order of $ecos(2\phi)$. The equation derived from the constant terms is, 

\begin{equation}  \begin{split}
\label{eqn}
& 4 \left( \frac{1}{4} - \delta^2 \right) e^2 f_o^2
-8 \left( \frac{1}{2} + \delta \right) ^2 e^2 \Omega_b f_o\\
& - 4 e e_b \left( \frac{1}{2} + \delta \right)
q_b = -q
 + \left( f_o + \Omega_b \right)^2 . 
\end{split} \end{equation}

\noindent where we use the following, simplifying change of variables,

\begin{equation}
\label{eqn0}
q = c p^{-2(1+\delta)}, \ \ q_b = c p^{-2(1+\delta_b)}.
\end{equation}

\noindent Equation \eqref{eqn} can be viewed as a quadratic in the coefficient $f_o$.  Then the equation derived from first order terms, i.e., terms in $ecos(2\phi)$, can be solved for the coefficient $f_1$. It is,

\begin{equation}  \begin{split}
\label{eqn1}
& 2 \left( f_o + \Omega_b \right) f_1 = 
4  \left( \frac{1}{2} + \delta \right) f_o^2
+ 2 \left( \frac{1}{2} + \delta \right)^2 
q\\
& + 2 \delta_b \frac{e_b}{e} q_b
+ \left( \frac{1}{2} + \delta \right)
\left( f_o + \Omega_b \right)^2. 
\end{split} \end{equation}

\noindent The equation in terms in $e^2 cos^2(2\phi)$, can be solved for the coefficient $f_2,$

\begin{equation}  \begin{split}
\label{eqn2}
& 2 \frac{\left( f_o + \Omega_b \right)}{\left( \frac{1}{2} + \delta \right)}
 f_2 = 8f_o^2 + 10f_o f_1 + 8\left( \frac{1}{2} + \delta \right) 
 f_o \Omega_b\\
 & + 2f_1 \Omega_b - \frac{1}{2} \left( \frac{3}{2} + \delta \right)
 \left( f_o + \Omega_b \right)^2
- \frac{f_1^2}{\left( \frac{1}{2} + \delta \right)} \\
& + \left( \frac{1}{2} + \delta \right)
\left[ 2 \left( \frac{1}{2} + \delta \right)^2 - 1 \right] q
 + 4 \left[ 1 + \frac{1}{2}\delta_b + \delta_b^2 \right]
\frac{e_b}{e} q_b. 
\end{split} \end{equation}

\noindent This completes the set of three equations for the $f_i$ coefficients. However, we still need to use the azimuthal equation \eqref{eqee} to solve for the p-ellipse variables, $p$ and $e$.

To begin, equation \eqref{eqm} can be differentiated to obtain an expression for the second derivative, $\ddot{\phi}$, which is, 

\begin{equation}
\label{eqo}
\ddot{\phi} \simeq 
-2 f_o f_1 e\ sin(2\phi)
-2 \left( f_1^2 + 2 f_o f_2 \right) 
e^2 sin(2\phi) cos(2\phi).
\end{equation}

\noindent This equation and equations \eqref{eqg}, \eqref{eqh}, \eqref{eql} and \eqref{eqm} can then be substituted into the azimuthal equation of motion \eqref{eqee} to obtain the perturbation constraint. In this azimuthal equation we retain only terms of first and second order in $e$, and after cancellation of a common factor of $e\ sin(e\phi)$, these appear as terms of zeroth and first order. This is confusing since the radial equation has true zeroth order terms giving the balance of gravitational and centrifugal forces when $e, e_b = 0$. Thus, we will continue to refer to these azimuthal equation terms as the first and second order conditions. The first order condition reduces to the following,

\begin{equation}
\label{eqp}
f_o f_1 = 2 \left(\frac{1}{2} + \delta\right) f_o 
\left( f_o + \Omega_b \right) 
+ \frac{e_b}{e} q_b.
\end{equation}

\noindent The second order equation ($e cos(2\phi)$ terms in the azimuthal equation) is, 

\begin{equation}  \begin{split}
\label{eqq}
& f_1^2 + 2f_o f_2 = 2 \left( \frac{1}{2} + \delta \right) \\
& \times \left[ -\left( f_o + \Omega_b \right) f_o
+ \left( 2f_o + \Omega_b \right) f_1
+ \left( 1 + \delta_b \right) \frac{e_b}{e} q_b \right].
\end{split} \end{equation}

\noindent This completes the set of coefficient equations derived from the equations of motion in this resonant case. The equations are linear in the variables $q, q_b e_b/e$, and of quadratic order or less in the $f_i$ factors. {\it Thus, to second order in any disc region with fixed values of $\delta, \delta_b$, there are generally zero to two resonant p-ellipse orbits.} This important result is evidently due to the fact that the ratio of epicyclic/precession frequency to circular orbit frequency is a constant in power-law potentials. We will consider specific solutions in the following section.

\subsection{Two frequency case}

In this subsection we consider second order solutions to the equations of motion (eqs. \eqref{eqee}) of the form of equation \eqref{eqfa}. The perturbation expansion procedure is essentially the same as that of the previous subsection. For brevity, we will not give the equations analogous to equations \eqref{eqg} - \eqref{eql} above. In this case the angular velocity expansion form, analogous to equation \eqref{eqm}, is, 

\begin{equation} \begin{split}
\label{eqm1}
&\dot{\phi} \simeq f_o + f_1 e cos(m\phi) + f_2 e cos(2\phi) \\
&+ f_3 e^2 cos^2 (m\phi) + f_4 e^2 cos^2 (2\phi) + f_5 e^2 cos^2 ((2-m)\phi).
\end{split} \end{equation}

Then, the coefficient equations deriving from the radial equation are analogous to equations \eqref{eqn} - \eqref{eqn2}, except there are now six of them. The first is the equation derived from the constant terms,

\begin{equation}  \begin{split}
\label{eqq1}
& \left( \frac{1}{2} + \delta \right)  \left( \frac{3}{2} + \delta \right) 
\left( m^2 + 4c_2^2 \right) e^2 f_o^2
- 4 \left( \frac{1}{2} + \delta \right) c_2 e e_b q_b\\
&-2 \left( \frac{1}{2} + \delta \right)^2 f_o 
\left( \Omega_b + f_o \right) \left( m^2 + 4c_2^2 \right) e^2
= -q + \left( f_o + \Omega_b \right)^2 ,
\end{split} \end{equation}

\noindent The equation from the $e\ cos(m\phi)$ terms is,

\begin{equation}  \begin{split}
\label{eqq2}
& 2 \left( f_o + \Omega_b \right) f_1 = 
\left( \frac{1}{2} + \delta \right) m^2 f_o^2
+ 2 \left( \frac{1}{2} + \delta \right)^2  q\\
& + 2 \delta_b \frac{e_b}{e} q_b
+ \left( \frac{1}{2} + \delta \right)
\left( f_o + \Omega_b \right)^2. 
\end{split} \end{equation}

\noindent The equation from the $e\ cos(2\phi)$ terms is,

\begin{equation}  \begin{split}
\label{eqq3}
& 2 \left( f_o + \Omega_b \right) f_2 = 
4 \left( \frac{1}{2} + \delta \right) c_2 f_o^2
+ 2 \left( \frac{1}{2} + \delta \right)^2  c_2 q\\
& + 2 \delta_b \frac{e_b}{e} q_b
+ \left( \frac{1}{2} + \delta \right) c_2
\left( f_o + \Omega_b \right)^2. 
\end{split} \end{equation}

\noindent The first second order equation from the $e^2\ cos^2(m\phi)$ terms is,

\begin{equation}  \begin{split}
\label{eqq4}
& 2 \left( f_o + \Omega_b \right) f_3 = 
2 \left( \frac{1}{2} + \delta \right) m^2 f_o f_1 \\
& - 2 \left( \frac{1}{2} + \delta \right)  
\left( \frac{3}{2} + \delta \right)m^2 f_o^2
 + 2 \left( \frac{1}{2} + \delta \right)^2 m^2
 f_o \left( f_o + \Omega_b \right) \\
& + \left( \frac{1}{2} + \delta \right)^2
\left[ 2 \left( \frac{1}{2} + \delta \right)^2 - 1 \right] q
- f_1^2\\
& - \frac{1}{2} \left( \frac{1}{2} + \delta \right)  
\left( \frac{3}{2} + \delta \right) \left( f_o + \Omega_b \right)^2
 + 2 \left( \frac{1}{2} + \delta \right)
f_1 \left( f_o + \Omega_b \right). 
\end{split} \end{equation}

\noindent The equation from the $e^2\ cos^2(2\phi)$ terms is,

\begin{equation}  \begin{split}
\label{eqq5}
& \frac{2}{c_2} \left( f_o + \Omega_b \right) f_4 = 
\left( \frac{1}{2} + \delta \right) f_2
 \left( 9 f_o + \Omega_b \right) \\
& - 8 \left( \frac{1}{2} + \delta \right)  
\left( \frac{3}{2} + \delta \right) {c_2} {f_o}^2
 + 8 \left( \frac{1}{2} + \delta \right)^2 {c_2}
 f_o \left( f_o + \Omega_b \right) \\
& + 4 \left( \frac{1}{2} + \delta \right)
\left( \frac{3}{2} + \delta \right) \frac{e_b}{e} q_b\\
&+ \left( \frac{1}{2} + \delta \right)^2
\left[ 2 \left( \frac{1}{2} + \delta \right)^2 - 1 \right] {c_2} q\\
& - \frac{1}{2} \left( \frac{1}{2} + \delta \right)  
\left( \frac{3}{2} + \delta \right) {c_2} \left( f_o + \Omega_b \right)^2.
\end{split} \end{equation}

\noindent And the equation from the $e^2\ cos^2((2-m)\phi)$ terms is,

\begin{equation}  \begin{split}
\label{eqq6}
& 2 \left( f_o + \Omega_b \right) f_5 = \\
&- \left( \frac{1}{2} + \delta \right) 
\left[ \left( \frac{3}{2} + \delta \right) \frac{c_2}{2} - c_x \right]  
\left( f_o^2 + \left(f_o + \Omega_b \right)^2 \right) \\
& - \frac{m}{2} \left( \frac{1}{2} + \delta \right)  
\left[ 8 \left( \frac{1}{2} + \delta \right) c_2 f_o
\left( f_o + \Omega_b \right) + 2 \frac{e_b}{e} q_b \right]
\\
&+ \frac{1}{2} \left( \frac{1}{2} + \delta \right)^2
\left[ \frac{c_2}{2} \left( \left( \frac{1}{2} + \delta \right)^2 - 1 \right) + c_x \right] 
q\\
& + \frac{\delta_b}{2}  \left( \frac{1}{2} + \delta \right)
\left( \frac{1}{2} + \delta_b \right) \frac{e_b}{e} q_b\\
& + \left( \frac{1}{2} + \delta \right)  
\left( f_2 + c_2 f_1 \right) \left( f_o + \Omega_b \right) 
+ f_1 f_2.
\end{split} \end{equation}

\noindent As in the previous case, most of these equations can be used to obtain values of the $f_i$ coefficients in equation \eqref{eqm1}. To proceed, we differentiate the quantity $\dot{\phi}^2$, derived from equation \eqref{eqm1} to get, 

\begin{equation}  \begin{split}
\label{eqo1}
&-\ddot{\phi} \simeq 
f_o \left[ mf_1 e\ sin(m\phi)
+ 2f_2 e\ sin(2\phi) \right]
\\
& + \left( f_1^2 + 2 f_o f_3 \right) 
m e^2 sin(m\phi) cos(m\phi) 
\\
& + 2 \left( f_2^2 + 2 f_o f_4 \right) 
e^2 sin(2\phi) cos(2\phi)
\\
& + \frac{2-m}{2} \left( f_1 f_2 + 2 f_o f_5 \right) 
e^2 sin((2-m)\phi) cos((2-m)\phi).
\end{split} \end{equation}

This equation can be used to eliminate the $\ddot{\phi}$ term in the angular equation of motion, as in the previous case. Then we obtain five coefficient equations by gathering like terms in this equation. The first of these is obtained from the $me\ sin(m\phi)$ terms,  
 
\begin{equation}
\label{eqp1}
f_1 = 2 \left(\frac{1}{2} + \delta\right) 
\left( f_o + \Omega_b \right).
\end{equation}

\noindent The equation from the $2e\ sin(2\phi)$ terms is, 
 
\begin{equation}
\label{eqp2}
f_o f_2 = 2 \left(\frac{1}{2} + \delta\right) 
c_2 f_o \left( f_o + \Omega_b \right)
+ \frac{e_b}{e} q_b.
\end{equation}

\noindent The equation from the $me^2\ sin(m\phi)cos(m\phi)$ terms is, 

\begin{equation}
\label{eqp3}
2f_o f_3 = 2 \left(\frac{1}{2} + \delta\right) 
 \left( f_o + \Omega_b \right)
 \left( 2f_1 - f_o \right) - f_1^2.
 \end{equation}

\noindent The equation from the $2e^2\ sin(m\phi)cos(m\phi)$ terms is, 

\begin{equation}  \begin{split}
\label{eqp4}
&2f_o f_4 = 2 \left(\frac{1}{2} + \delta\right) c_2^2
 \left( f_o + \Omega_b \right)
 \left( 2f_2 - f_o \right) - f_2^2 \\
 & + \frac{c_2}{2} \left(\frac{1}{2} + \delta \right) 
 \left(\frac{1}{2} + \delta_b \right) \frac{e_b}{e} q_b.
 \end{split}   \end{equation}

\noindent And the equation from the $e^2\ sin((2-m)\phi)$ terms is, 

\begin{equation}  \begin{split}
\label{eqp5}
&(2-m) f_o f_5 = \\
&2 \left(\frac{1}{2} + \delta\right) 
\left( f_o + \Omega_b \right) \left[
-mf_2 + 2 c_2 f_1 
 + (2-m) \left( -\frac{c_2}{2} + c_x  \right) f_o \right]
 \\
& - \frac{2-m}{2} f_1 f_2
 + c_2 \left(\frac{1}{2} + \delta \right) 
 \left(\frac{1}{2} + \delta_b \right) \frac{e_b}{e} q_b.
 \end{split}   \end{equation}
 
 These five equations from the azimuthal equation, together with the six from the radial equation (eqs. \eqref{eqq1} - \eqref{eqq6}) complete the set needed to solve for the parameters $f_o - f_5, m, p, e, c_2$, and $c_x$ of the approximate solution given by equations \eqref{eqfa} and \eqref{eqm1}.  
 
\section{Approximate loop orbit solutions}

In this section we explore solutions to the perturbation equations of Sec. 3.1 based on the simple p-ellipse of equation \eqref{eqf}. These solutions may be parents of families of orbits in non-self-gravitating galaxy bars, which are driven by an external potential, as discussed below. The external potential may due to a prolonged tidal perturbation, or a bar-like dark halo. 

\subsection{A very simple special case}

Solutions to the single frequency cases discussed in Sec. 3.1 are determined by the coefficient equations \eqref{eqn}, \eqref{eqn1}, \eqref{eqn2}, \eqref{eqp}, and \eqref{eqq}. We note that the sum $f_o + \Omega_b$ is a common term in these equations, and in the case where $f_o = -\Omega_b$ the equations are simplify significantly. This is the case we consider in this subsection. We note that this case has no special physical significance. The factor $f_o$ is the mean rotation frequency of the star in the pattern frame, and there is no obvious reason for it to equal the opposite of the pattern frequency. 

However, this simple case suggests a simple analytic solution strategy, which can be generalized. This strategy makes use of the fact that if we assume the value of one of the unknowns ($f_o$), then we can treat the factor $(e_b/e)q_b$ as an unknown variable, even though it is actually a combination of the variables $e, p$, and the presumably known potential amplitude $e_b$. We are inverting the direct problem of finding to $f_o$ to ask what value of $e_b$ is needed to get the assumed value of $f_o$.

Then, the solution is obtained via the following procedure. First, use equations \eqref{eqp} and \eqref{eqn1} to eliminate the variables $f_1$ and $(e_b/e)q_b$, respectively, from equation \eqref{eqn2}. The latter is a quadratic that gives $q$ in terms of $\delta, \delta_b$, and $f_o$. The value of $q$ yields the value of $p$ and $q_b$, and the remaining solution parameters are obtained directly from the other equations. 

 We require the solution for $q$ to be a positive, real number. In this special case, the quadratic solutions are imaginary or negative for a large range of parameter values. Even when there is a real, positive solution (or two), other physical constraints must be satisfied, e.g., $0 \le e \ll 1$. A range of parameter values have been explored, and relatively few physical solutions have been found in this case. 
 
 \subsection{More general closed loop orbits}
 
 Fortunately, when we deviate from the special case of the previous subsection ($f_o = -\Omega_b$), we find more physical solutions to the coefficient equations, i.e., closed loop orbits. We consider several examples in this subsection. Nonetheless, we will follow the same procedure for solving the coefficient equations as in the previous subsection. Specifically, we will adopt a value of the pattern speed, $\Omega_b$, and a value of the mean orbital speed of the star, $f_o$, as some multiple of the former. In principle, we could adopt values of the bar parameters,  $\Omega_b, e_b$ (and $\delta_b$), and then solve for the solution parameters $e, p, f_i$. However, as discussed above the solution is easier to obtain if we assume a value of $f_o$ and derive the corresponding value of $e_b$. In the following two subsections we consider some specific examples of physically relevant solutions. 
 
 \subsubsection{Slowly rising rotation curve examples}

The first sequence of examples has relatively slowly rising rotation curves appropriate to the inner part of a galaxy disc in both the symmetric part of the potential (with $\delta = -0.3$) and the asymmetric part (with $\delta_b = -0.2$). We also adopt the (arbitrary) value of $\Omega_b = 0.315$, and consider a range of values of $f_o$ and the ratio $n_b = -\Omega_b/f_o$. 

For a first example we take $f_o = -0.3$ and $n_b = 1.05$; the solution of the coefficient equations then yields: $p = 9.56, e = 0.61$, and $e_b = 5.68$. All of these solution values are relatively large, so we might not expect the perturbation approximation to be very accurate in this case. Figure 1 compares the p-ellipse approximation with these parameters to a numerically integrated orbit with the same initial conditions in the pattern frame. I.e., the initial conditions are $\phi = 0, dr/dt = 0$, with $r$ given by the p-ellipse equation and $d\phi/dt$ given by a value like that of equation \eqref{eqm}, with $c = 1$. It is apparent that the two orbits are very similar.  A small subharmonic modulation (four times the fundamental period) is visible in the numerical orbit in the lower panel, which presages a trend we will see more of below. This modulation is also responsible for the finite thickness of the numerical orbit curve in Fig. 1.

While the fit of the analytic to the numerical curve in Fig. 1 is impressive, there is an important caveat. In the previous paragraph, the initial angular velocity used in the numerical orbit was described as ``like that of equation \eqref{eqm}.'' Although the value predicted by equation \eqref{eqm} is $\dot{\phi} = -0.49$, while the value that yields the good fit is \eqref{eqm} is $\dot{\phi} = -0.87$. Thus, the analytic equation for $\dot{\phi}$ does not yield an accurate approximation for the values of $e$ as large as the present example. This is understandable, since in the present case the analytic prediction is that the terms of equation \eqref{eqm} are $(f_o, ef_1, 0.5e^2f_2) = (-0.3, -0.51, 0.16)$. Clearly, the series on the right-hand-side of equation \eqref{eqm} is not converging rapidly with the relatively large value of $e$. This slow $\dot{\phi}$ convergence is a limitation on the p-ellipse approximation, but it is a predictable consequence of large values of $e$. Note that the more accurate value for the numerical orbit was found by trial and error, and so too in most of the examples below. 

Nonetheless, this first example shows that a p-ellipse approximation, developed for orbits in symmetric potentials, can also fit rather flattened orbits in two-part, asymmetric potentials quite well. One major difference between the adopted p-ellipse solution and those used for symmetric potentials (see \cite{st06}) is that we have changed the frequency ratio to a fixed resonant value, rather than the value representing the precession of the given symmetric potential. 

\begin{figure}
\centerline{
\includegraphics[scale=0.5]{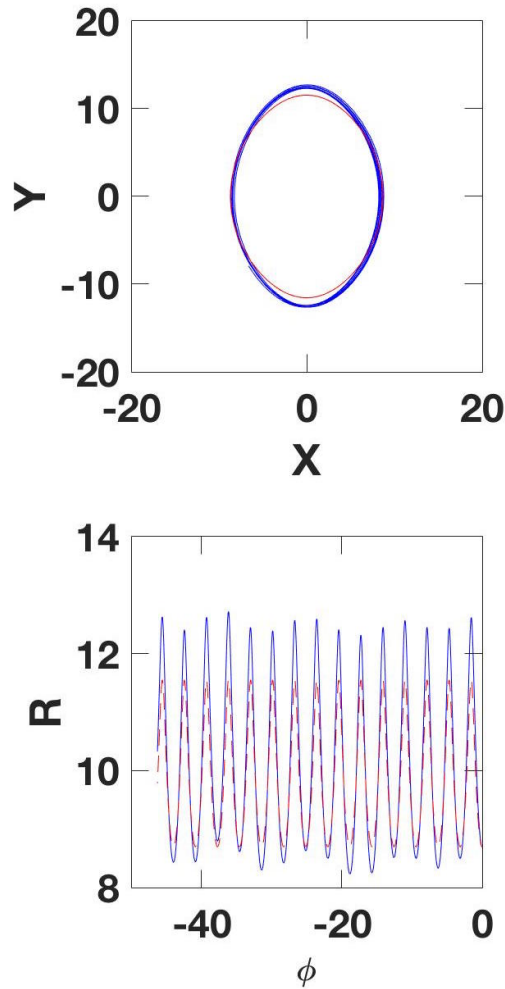}}
\caption{A sample orbit determined by the parameter value $f_o = -0.3$ and in the potential specified by the values $\delta = -0.3, \delta_b = -0.2$ and pattern speed $\Omega_b = 0.315$, in the dimensionless units. In both panels the blue solid curve is the result of numerically integrating the equations of motion with the given initial conditions (see text for details), and the red dashed curve is the p-ellipse approximation. The upper panel is the view onto the disc in the pattern frame; the lower panel shows radius versus azimuthal advance, which is negative in this case.}
\end{figure}

A second orbital example, specified by $f_o = -0.3029, n_b = 1.04$ and the same pattern frequency ($\Omega_b = 0.315$), and with derived analytic values of $p = 5.0076, e = 0.9542$, and $e_b = 3.2069$ is shown in Fig. 2. In this high eccentricity case, the predicted value of $\dot{\phi}$ was -0.32, and the fitted value, -1.19. Although the value of $f_o$ is only slightly changed from the previous orbit, this orbit is much smaller and more elongated. In fact, physical orbit solutions can generally only be found over a small range of $f_o$ values. 

Two differences from the previous example are evident in the top panel of Fig. 2. First, the fit is not as good. This is not surprising given the high eccentricity. (Note, that the flattening at a given p-ellipse eccentricity differs from that of a simple ellipse, see \citet{st06}.) Secondly, the numerical orbit is thicker, a result of stronger subharmonic modulation, which is evident in the lower panel of Fig. 2. 

\begin{figure}
\centerline{
\includegraphics[scale=0.65]{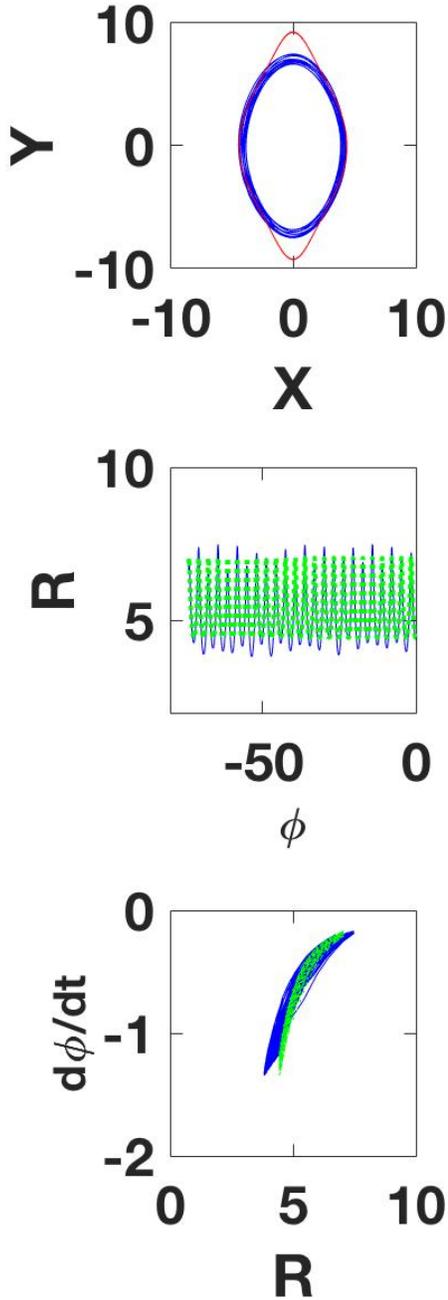}}
\caption{Like Fig. 1, but for the orbit determined by the parameter value $f_o = -0.3029$, again in the $\delta = -0.3,\ \delta_b = -0.2$ potential with pattern speed $\Omega_b = 0.315$. The blue solid curve shows the numerically integrated orbit in both panels. The red dashed curve in the upper panel is the analytic orbit. The green dotted curve in the lower panels is the analytic curve, but corrected as described in the text. The lowest panel shows the azimuthal velocity in the pattern frame as a function of radius. Note that the azimuthal velocity is negative, and the speed is higher at small radii.}
\end{figure}

As \citet{ly10} found for elliptical orbits, and \citet[see Eq. 12]{st15a} confirmed for p-ellipses, a given orbit can be approximated much more accurately by using the eccentricity derived from the values of its inner and outer radii, rather than that derived as above. In the present case, once we have found a closed numerical orbit that best agrees with the p-ellipse approximation, we can use its extremal radii to get a better estimate of the eccentricity (here $e = 0.825$). This latter approach is used in the lower panel of Fig. 2, and the result is an excellent fit despite the high eccentricity (except for the subharmonic modulation). 

The difference between these two examples was a slight increase of the parameter $f_o$ in the second case. If we increase $f_o$ further (but still slightly) we find the trends described above continue. I.e., the numerical orbits tend to get a little smaller and more eccentric, but the p-ellipse approximation tends to exaggerate the eccentricity by greater amounts, unless corrected as just described. 

If instead we decrease the (negative) value of $f_o$ from the value used in the first example, then the trends reverse. I.e., we get bigger orbits, that are less eccentric, but also tend have higher subharmonic modulation; they are thicker. This last trend continues up to the point that the (numerical) orbits change their shape altogether. 

A third example shows the nature of this shape change with the present potentials. This case, shown in Fig. 3, has $f_o = -0.2990, n_b = 1.0535$ and the same pattern frequency as the previous examples. The derived analytic values are: $p = 15.79, e = 0.4301$, and $e_b = 8.8691$. The predicted value of $\dot{\phi}$ was -0.4973, and the fitted value, -0.7142.  In this case the resonant orbit has three distinct loops, which cannot be fit by a single loop p-ellipse. However, the p-ellipse approximation appears as a low-radius boundary, and provides a fit only in the vicinity of the lowest radial excursion. That is a general result for such multi-loop orbits. The basic peak and trough phases of the numerical model are also captured by the analytic model. This example suggests that the subharmonic component is becoming much stronger as we decrease the value of $f_o$. 

Slight variations in the initial angular velocity yields a family of orbits that are similar, but with thicker loops. With enough initial velocity deviation the three-loop form disappears, and the orbits are better described as a filled annulus between the p-ellipse and the outer loop. It is also true that initial velocity variations produce thicker loops in the previous examples. Thus, each closed, resonant orbit has a family of such offspring, extending over a finite interval of the initial value of $\dot{\phi}$. 

\begin{figure}
\centerline{
\includegraphics[scale=0.5]{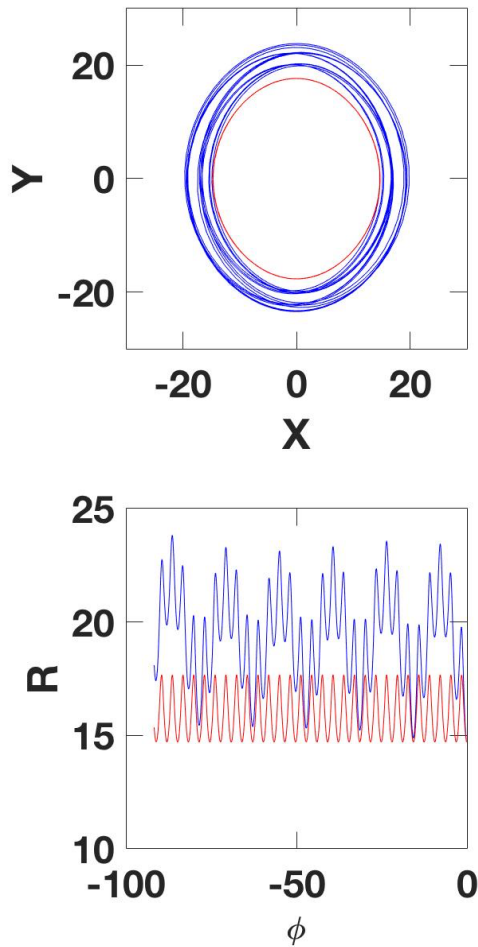}}
\caption{Like Fig. 1, but for the orbit determined by the parameter value $f_o = -0.2990$, again in the $\delta = -0.3,\ \delta_b = -0.2$ potential with pattern speed $\Omega_b = 0.315$. The blue solid curve shows the numerically integrated orbit, and the red dashed curve in the upper panel is the analytic orbit in both panels. The sub harmonic modulation is clear in the lower panel.}
\end{figure}

\begin{figure}
\centerline{
\includegraphics[scale=0.5]{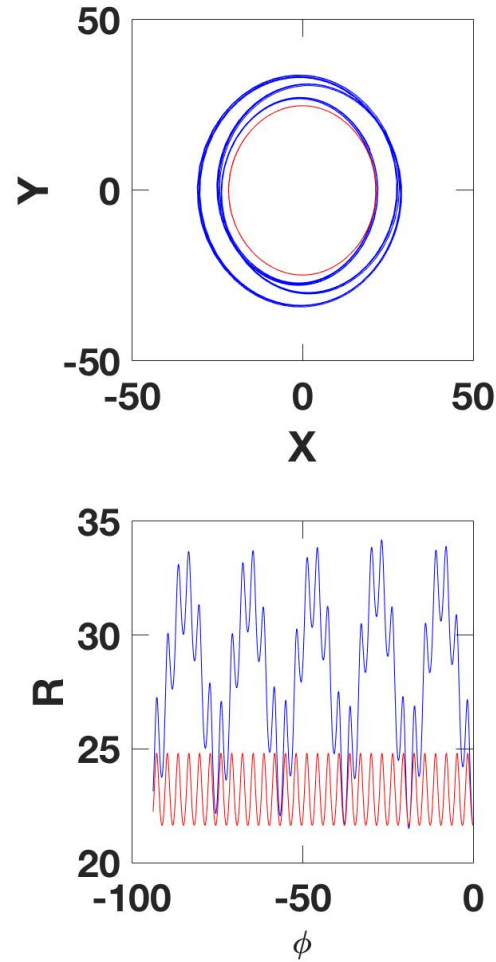}}
\caption{Like Figs. 1, and 3, but for the orbit with $f_o = -0.2986$. The orbit is similar to that in Fig. 3, but larger and broader. The lower panel shows that the subharmonic frequency is becoming dominant.}
\end{figure}

Figure 4 shows a fourth example, with a still larger value of $f_o = -0.2986 (n_b = 1.0549)$, and the same pattern frequency. In this case, the derived analytic values are: $p = 22.90, e = 0.3305$, and $e_b = 12.32$. The predicted value of $\dot{\phi}$ was -0.4788, and the fitted value, -0.6120. Again, we see a three loop pattern, with the analytic orbit serving as a low radius boundary to the numerical orbit. The subharmonic frequency is longer relative to the analytic period than in the previous example, and their combined width is greater. It is clear that the centre of each loop is offset from the origin, in alternating directions along the x-axis. If we increase the value of $f_o$ to -0.2984 the analytic approximation breaks down entirely, yielding negative values of $p$, the radial scale. Between the current value of $f_o$ and that critical value, the overall orbit size and the spread between loops continues to get larger. The analytic eccentricity, and that of the loops, decreases. The offset of loop centres also increases. 

Given the large values of $e_b$ in these last two or three examples, a quantity assumed to be of order $e$ in the perturbation expansion, it is not surprising that the analytic approximation breaks down. It is more surprising that it continues to provide some information as it breaks down. 

In sum, all of these sample orbits have the same pattern speed, so they, and their less regular offspring could be combined to make a model bar. However, this model would require increasing bar strength ($e_b$) with increasing radius. We might not expect the bar to extend beyond the radius where the orbit breaks into multiple loops, because the pattern becomes less distinct and more circular. Gaseous components would experience dissipation and circularization at loop meeting points. This radius apparently differs from the co-rotation radius usually thought to give the outer extent of bars. 

Alternately, if the needed bar strength occurs only over a limited range of radii, then a hollow annular stellar bar would be possible. In this latter case, if the value of $e_b$ changed with time the bar structure would evolve. E. g., it could grow larger and wider as $e_b$ increases.  

\subsubsection{More steeply rising rotation curve examples}

\begin{figure}
\centerline{
\includegraphics[scale=0.5]{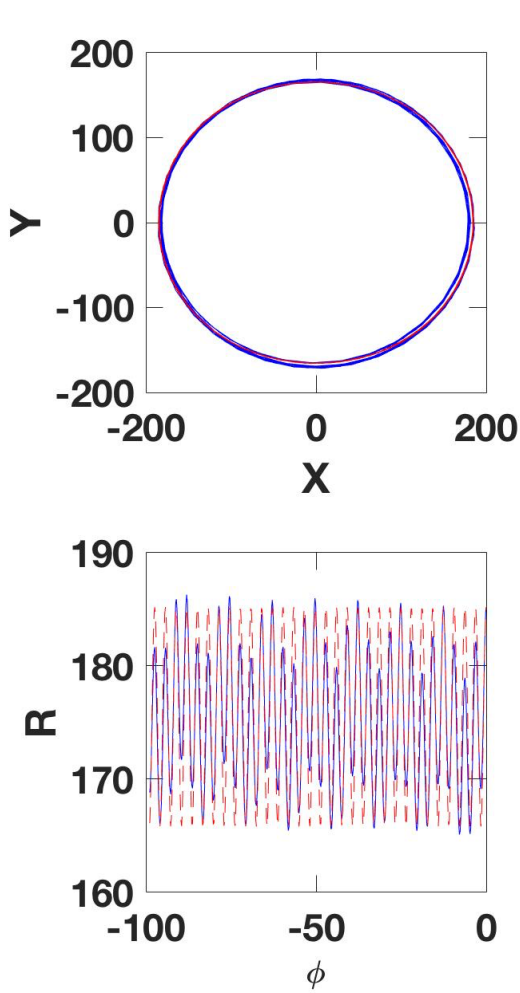}}
\caption{A sample orbit determined by the parameter value $f_o = -0.77 (f_1 = 1.025, f_2 = 0.464)$ and in the potential specified by the values $\delta = -0.8, \delta_b = -0.7$ and pattern speed $\Omega_b = 1.05$, in the dimensionless units. In both panels the blue solid curve is the numerically integrated orbit, and the red dashed curve is the p-ellipse approximation. The upper panel is the view onto the disc in the pattern frame; the lower panel shows radius versus the negative azimuthal advance. Note the large size, and near circularity of the orbit.}
\end{figure}

\begin{figure}
\centerline{
\includegraphics[scale=0.64]{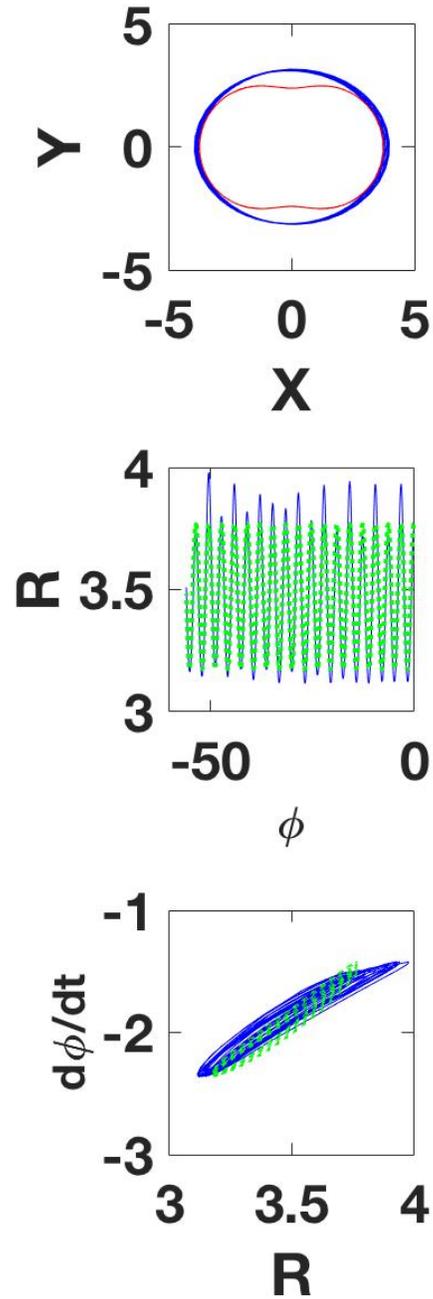}}
\caption{Like Fig. 5, but for the orbit with $f_o = -0.715 (f_1 = 1.007, f_2 = 0.540)$ in the same potential, with the same pattern speed. The blue solid curve shows the numerically integrated orbit in both panels. The red dashed curve in the upper panel is the analytic orbit, which does not yield a good fit in this case. The green dotted curve in the lower panels is the analytic curve, corrected with a lower eccentricity and as further described in the text.}
\end{figure}

In this subsection we consider a second set of examples drawn from a potential with a more steeply rising rotation curve, indeed close to a solid-body potential. Specifically, we take $\delta = -0.8$ in the symmetric part of the potential and $\delta_b = -0.7$ in the asymmetric part. The case with $\delta = -0.5$ is a singular one, where the perturbation approach above breaks down. The character of the orbits is rather different for $\delta$ values on either side of this critical value. 

One difference is that the resonant, closed orbits generally only exist at higher pattern speeds than in the previous case. For the examples in this section we take $\Omega_b = 1.05$. We will consider three orbits in this bar pattern.

The first, shown in Fig. 5, is a large, but low eccentricity one, specified by the parameter value $f_o = -0.77$. The numerical orbit has a small, but finite width, and some subharmonic modulation is visible in the lower panel.  Not surprisingly, the analytic fit to this nearly circular orbit is very good. 

We skip over a range of large, low eccentricity orbits to the much smaller, and visibly flatter one in Fig. 6. Though flatter, this orbit does not pinch inward like the analytic approximation. The fit is not good, but we can improve it using the maximum and minimum radii of the numerical orbit to derive new values of $e$ and $p$, as described for the Fig. 2 orbit. The revised fit is good and shown in the lower panels as a green dotted curve, which basically overwrites the blue numerical curve in the region of overlap. The thickness of the numerical orbit is larger relative to its mean radius than that of the orbit of Fig. 5, but it is still small.

\begin{figure}
\centerline{
\includegraphics[scale=0.5]{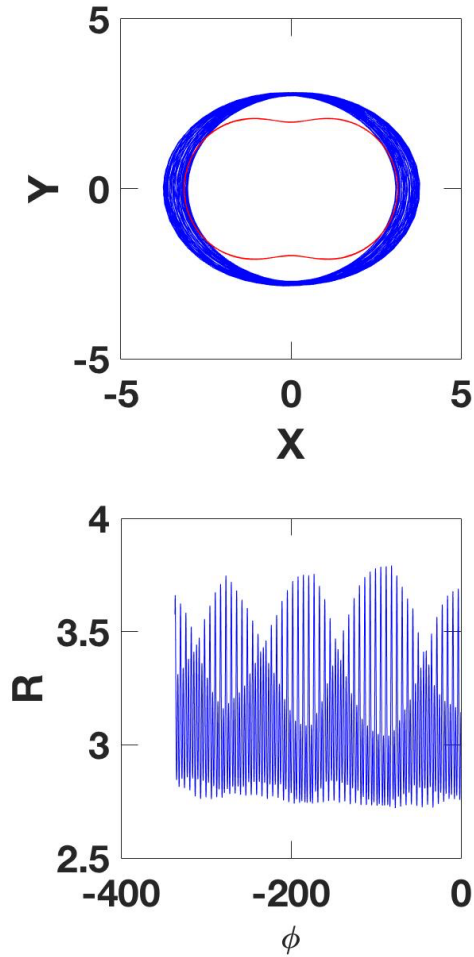}}
\caption{Like Fig. 6, but for the orbit with $f_o = -0.71$ in the same potential, with the same pattern speed. In the lower panel the analytic orbit has been omitted for clarity. The subharmonic pattern dominates.}
\end{figure}

The value of the parameter $f_o$ is changed only slightly between the cases shown in Figs. 6 and 7, but the effect is significant. Specifically, the orbit is beginning to get much thicker (and this trend accelerates for yet larger values of $f_o$). The fit of the analytic orbit is worse than in Fig. 6. In this orbit, and others not shown with lower values of $f_o$, the analytic curve approximates a portion of the inner boundary of the numerical orbit (as in Figs. 3,4). While the orbits get wider as $f_o$ is decreased, their inner boundary gets smaller and flatter only slowly. The lower panel of Fig. 7 shows that the subharmonic modulation is becoming dominant as in the more extreme examples of the previous subsection. In fact, two subharmonics are visible, one at twice the basic frequency, as well as the lower frequency one.

\subsection{Generalizations}

It is interesting to compare the orbit sequences of the last two subsections. The flattest orbits are the smallest in both cases, with larger orbits becoming more nearly circular. In none of the cases above are the orbits even close to the flatness apparent in some observed bars. Interestingly, this statement does not apply to the analytic bars of the case in Sec. 4.2.1. Fig. 2 hints at how these can get much longer and flatter than the corresponding 'true' numerical orbit. This along with other results described below and in the literature, suggests that if closed orbits play a significant role in flat bars, they are not simple loops. If that is true, then it is also likely that shocks in the gas also play an important role in such bars. 

The two sequences above also share the characteristic that at one end of the range of allowable $f_o$ values the subharmonic modulation becomes very strong, and the relative width of the orbit grows as fast or faster than the relative change in the mean radius. The two sequences differ, however, in which end of the spectrum shows this phenomena. In the first sequence (Sec. 4.2.1), it occurs at the smaller values of $f_o$, where the orbits are large and more circular. In the second sequence (Sec. 4.2.2), it occurs at larger values of $f_o$ where the orbits are smaller and flatter. This behavior reversal seems to occur generally across the $\delta = -0.5$ singularity, based on additional cases not presented here. 

It should also be noted that there appears to be a large region in the $(\delta, \delta_b, \Omega_b)$ parameter space where simple closed loop orbits do not exist. The cases with low pattern speeds, and values of $\delta, \delta_b < -0.5$ have already been mentioned. This also seems to be true for falling rotation curves with values of $\delta, \delta_b > 0.0$. We have not explored this parameter space extensively, so these conclusions are preliminary. The conclusion that bars are more likely in regions with rising rotation curves, an extrapolation of these loop orbit results, does seem in accord with observational and modeling results, e.g., \citep{se14}. On the other hand, loop-like orbits are found numerically in falling rotation curve potentials in parameter regions where the perturbation approximation fails to give solutions. This will be explored in a sequel paper.

\section{Loop bars in general potentials}

In the previous section we considered examples of sets of closed loop orbits in symmetric and bar-like power-law potentials. In the case of a single value of the bar potential amplitude, $e_b$ in the disc, there is generally one such orbit, though zero to two closed loop orbits are allowed by the perturbation theory in Sec. 3.1. However, the numerical results of the last section indicate that one such orbit is the most common result, and this result extends to quite high eccentricity. If the magnitude of $e_b$ increases outward there can be a number of nested near-loop orbits, which could form the skeleton of a bar. 

However, the potentials of galaxy discs are not well described by a single power-law in radius, but rather have rising rotation curves in the inner parts and flat or falling rotation curves in the outer regions. In this section we consider another series of loop orbits in a potential with varying power-law indices, as an example of a potential approximated by a sum of power-laws. The main point of this example is to show that results like those of the previous section can be generalized to potentials that can be approximated by such variable power-law forms. 

\begin{figure}
\centerline{
\includegraphics[scale=0.42]{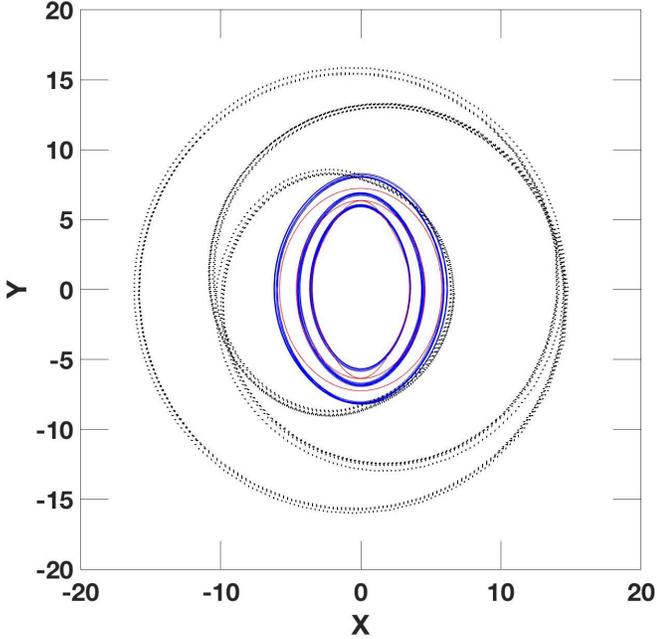}}
\caption{A sequence of four closed orbits with the same pattern speed, $\Omega_b = 0.315$. The three inner bar-like orbits derived by numerical integration are shown by thick, blue curves, and their analytic approximations are shown by thin, red curves. The fourth, outermost dotted orbit has consists of three loops, and no analytic approximation is shown. The initial conditions $(r_o, \dot{\phi_o})$ are: $(3.53, -1.23)$, $(4.60, -1.01)$, $(5.80, -0.837)$, $(6.68, -0.657)$.}
\end{figure}

Figure 8 shows four representative orbits in this example. The assumed rotation curve rises moderately steeply in the inner regions, but transitions to a flat rotation curve in the outer regions. The equipotentials are also assumed to be of about the same shape as the loop orbits. The values of the potential indices from the inner orbit outwards are, $\delta = \delta_b = (-0.25, -0.20, -0.15, 0.0)$. That is, we assume that local power-law potential approximations to the potential have these index values and that $\delta = \delta_b$ throughout. The value of the pattern speed is $\Omega_b = 0.315$. From the inside out, the values of $(p,e, e_b)$ for the analytic approximations to the inner three orbits are: $(4.11, 0.826, 1.97),$ $(5.19, 0.495, 2.53),$ $(6.36, 0.306, 3.31),$ $(7.03, 0.108, 4.89)$.

Despite their varying eccentricity, the three nested innermost orbits represent a clear bar structure. Between them and the outermost orbit there will be orbits with increasing thicknesses (and decreasing eccentricities), like those of Figs. 3, 4. As the potential index decreases towards $\delta = 0.0$ with increasing radius it becomes harder to get single loop orbits. Multiple (but not necessarily three) looped orbits become the rule at large radius. 

The one shown in Fig. 8 is particularly interesting because it represents a set derived from a small range of initial radii $r_o$ that have an inner loop, which partially overlaps a significant part of one of the simple inner loops. On the other hand, the outer loop of this orbit is nearly circular. A traversing the relevant part of the inner loop would, in some sense, look like it was pursuing a simple bar orbit. However, on the outer loop the star would look like it was on a circular orbit well outside the bar. Such orbits do not seem to have been studied in the bar literature (see e.g., the reviews of \citealt{at13, se14}), though orbits with small loops at their ends are common.  (However, in a recent paper \citet{ch17} find similar orbits in spherical potentials.) We note that it is a natural extension of the analytic approximation (via varying the value of $f_o$) that leads to them. As in the examples of the previous section, such closed, resonant orbits have a family of nearby (in the space of initial conditions) orbits that do not close. Evidently, their non-circularity would introduce a significant component of apparent velocity dispersion into the outer disc, beyond that of small bar perturbations of near circular orbits of large radii. They seem worthy of further study. 

In sum, though the example discussed in this section is ad hoc, it shows that a loop orbit skeleton of a model bar can be constructed in a potential more complex than that consisting of monotonic power-laws in both symmetric and asymmetric parts. The well studied Ferriers bars provide more examples (see \citealt{at92}). It also shows, as mentioned above, such a model bar effectively ends as the rotation curve becomes flat. This does not appear to be directly related to a co-rotation radius.

\section{The possibilities for self-consistent loop orbit bars} 

The gravitational potentials considered in the previous section were fixed, and presumably of external origin. We sought simple, closed, bar-like orbits in those potentials that could be analytically approximated with p-ellipses to second order in the eccentricity parameter. In this section we ask whether the bar potential itself could be constructed from such orbits? We will see that there are complications, and such bars are likely to be rare or non-existent, at least in two dimensions.

\subsection{Poisson equation and constraints}

To construct bar potentials from loop orbits we use the two-dimensional Poisson equation, which can be written in dimensionless units as,

\begin{equation} 
\label{eqt}
\triangledown^2 \Phi = \frac{1}{r} \frac{\partial}{\partial r}
\left( r \frac{\partial \Phi}{\partial r} \right) +
\frac{1}{r} \frac{\partial}{\partial \phi}
\left( \frac{1}{r} \frac{\partial \Phi}{\partial \phi} \right)
= 4\pi  \rho,
\end{equation}

\noindent where the scale factors are as in equation \eqref{eqd}, with addition of the following scales for the potential and density,

\begin{equation} 
\label{equ}
\Phi_\epsilon = c \frac{\epsilon^2}{\tau^2},\ \
\rho_\epsilon = \frac{M_\epsilon}{\epsilon^3}.
\end{equation}

\noindent  However, in the limit that bar does not have a significant effect on the halo potential, these two parts of the potential decouple, and the Poisson equation above can be assumed to describe the bar alone. Then we substitute the asymmetric potential term from equation  \eqref{eqb} to get the following expression for the mass density, 

\begin{equation} 
\label{eqv}
4\pi  \rho = \frac{4ce_b}{r^{2(1+\delta_b)}}
\left( 1 - \delta_b^2 \right)cos(2\phi),
\end{equation}

For this simple example, we assume a stationary bar, $\Omega_b = 0.0$.

We assume that we can construct the density field of the bar with an appropriate radial distribution with nested p-ellipse orbits (in the rotating frame). Then we can use the p-ellipse equation to eliminate factors of $r$ in the above. To second order, the expression for $\rho$ becomes, 

\begin{equation} 
\label{eqw}
\rho = a_o\ ecos(2\phi) \left[1 + a_1 e cos(2\phi) \right],
\end{equation}

\noindent with, 

\begin{equation}
\label{eqx}
a_o = \frac{c(1 - \delta_b^2)}{\pi p^{2(1+\delta_b)}}
\frac{e_b}{e}, \ \ 
a_1 = 2(1+\delta_b)\left( \frac{1}{2}+\delta \right).
\end{equation}

\noindent If $e$ varies slowly with radius, then the radial dependence of the density is given by the $p^{2(1+\delta_b)}$ term.

To relate the angular dependence of the density contributed by set of adjacent orbits to the azimuthal velocity on the orbits, we can assume that the more time a star spends on a given part of its p-ellipse orbit the greater its contribution to the density at that azimuth. Specifically, assume that the relative density change compared to that at $\phi = 0$ on a given part of the orbit equals the opposite of the relative azimuthal velocity change. I.e., 

\begin{equation} \begin{split}
\label{eqx1}
 \frac{\rho(\phi) - \rho_{\phi = 0}}{\rho_{\phi = 0}}
=  -\left( \frac{\dot{\phi}({\phi}) - \dot{\phi}_{\phi = 0}}{\dot{\phi}_{\phi = 0}} \right),
\\ or,\ 
\frac{\rho}{\rho_{\phi = 0}} =
2 - \frac{\dot{\phi}}{\dot{\phi}_{\phi = 0}}
\end{split} \end{equation}

\noindent Now we can substitute equation \eqref{eqm} for $\dot{\phi}$, equate our two expressions for the density (eqs. \eqref{eqw} and \eqref{eqx1}), and identify terms of common order in $e cos(2\phi)$. This yields, the following zeroth, first and second order equations, after some manipulation, 

\begin{equation} \begin{split}
\label{eqy}
f_o = 2ef_1 - 2e^2f_2,\\
\frac{c(1 - \delta_b^2)}{\pi p^{2(1+\delta_b)}} \frac{e_b}{e}
= \rho_{\phi = 0} \left( \frac{-f_1}{f_o + ef_1 + e^2f_2} \right),\\
(1+\delta_b)\left( \frac{1}{2}+\delta \right) = 
\frac{f_2}{2f_1}.
\end{split} \end{equation}

\noindent These equations provide strong additional constraints for self-gravitating loop bars. The kinematic orbits, discussed in previous sections, were not so constrained. Free parameters included $e_b$ (or $f_o$), $\delta_b$, $\delta$ and the pattern speed $\Omega_b$, though closed loops generally only existed for isolated values of $f_o$. As discussed in the next subsection, the extra constraints eliminate most of these solutions. 

\subsection{Self-consistent loop orbit bars?}

Already, in the non-self-gravitating cases above we found large areas of parameter space with no physical orbit solutions to the perturbation equations, and the lack of closed loops in the parameter space neighborhood was confirmed with numerical orbit integrations. With the additional constraints from the Poisson equation, the regions of parameter space with loop orbits seem to very small. This is evident just from the first of equations \eqref{eqy}, which provides another relation between the $f_i$ coefficients of the azimuthal velocity (and $e$). For example, we can follow the procedure of the previous sections to determine a value of $f_o$ iteratively for given values of $\delta_b$, $\delta$ and $\Omega_b$ that yields a loop orbit if it exists. Then the values of $f_1,\ f_2$ and the eccentricity parameter are also determined. However, the odds that these values incidentally satisfy the first of equations \eqref{eqy} will generally be very small. This consideration alone eliminates most of the loop solutions of the previous sections. 

The second of equations \eqref{eqy} can be viewed as an expression for the density variation along the $\phi = 0$ axis, and so, does not  constrain the solutions. 

The third equation of the set is constraining in several ways. The first is that since it gives a relation between $\delta$ and $\delta_b$, so one free parameter is eliminated. Loop orbits were found previously only for certain values of those two variables (for a given pattern speed), and those solutions which do not happen to also solve this third condition will be eliminated. This constraint is not as stringent as that imposed by the first equation, since a wide range of $\delta,\ \delta_b$ values produce loops. 

This third equation also provides some more detailed constraints. Recall that the physical range for the values of $\delta$ and $\delta_b$ is about $-1.0$ to $0.5$. The value of the factor $1 + \delta_b$ in that equation is never negative over this range of $\delta_b$ values. On the other hand, the values of $f_1$ and $f_2$ can have either sign. For example, in the case shown in Figs. 1-4 where $\delta,\ \delta_b > -0.5$, $f_1 <  0$ and $f_2 > 0$. With these values the left side of the third equation is positive and the right negative, so solutions like those of Figs. 1-4 cannot satisfy the Poisson constraints. We must have $\delta_b < -0.5$ when $f_2/f_1 < 0$. 

The orbits shown in Figs. 5-7 provide another example with $\delta,\ \delta_b < -0.5$. Here both $f_1$ and $f_2$ have positive values, so with these values the third equation is again violated, and more orbit solutions are precluded. Although we only have examples, not a rigorous proof, it appears that if the values of $\delta$ and $\delta_b$ are close to each other, then the Poisson constraint is violated. A third example is when $\delta_b < -0.5$. Then the factor $(1+\delta_b)$ will be small, and since the ratio $f_2/f_1$ is often of order unity, the factor $f_2/(2(1+\delta_b)f_1)$ is likely to be of order unity or larger. Yet if $\delta > -0.5$, then $(0.5 + \delta)$ is less than unity (unless the value of $\delta$ is near 0.5), so the Poisson constraint is not satisfied. (We have found cases like this, that yield kinematic loop orbits, but did not describe them above.) A more systematic analysis of the various cases could be done, but these several examples make the point that the third of equations \eqref{eqy} is very constraining. 

Thus, it we conjecture that a self-consistent bars can be constructed from loop orbits alone, in two dimensions, only in rare, or in no cases. Evidently, the orbits in self-consistent bars must be more complex in two dimensions. Moreover, it appears likely that the restrictions above would also apply to loop orbits with small librations; these are also not sufficiently complex. 

The restrictions above may, however, be loosened in a three-dimensional bar. Consider a very simple example of a cylindrical bar consisting of loops parallel to the disc plane, but not all lying within that plane. For example, suppose their vertical distribution was described by an exponential term, $e^{-|z|/z_o}$, in the density and potential. The z-derivatives of the three-dimensional Poisson equation would introduce $z_o^2$ terms in the last two constraint equations \eqref{eqy}. Then the last of these constraints could be viewed as an equation for this new parameter. If $z_o$ was a function of radius, then more parameters, describing this variation, would be introduced. These additional parameters would allow a broader range of solutions, and at least in principle, allow for cylindrical loop bars. We will not explore this topic further here.

\section {Ramifications for tidal and gaseous bars}

The possibility of generating bars in flyby galaxy collisions is natural because both the tidal force and the bar have the same basic symmetry. \citet{no87, no88} first investigated this with numerical hydrodynamical simulations (also see \citealt{ba91}). The simulations of \citet{ge90} demonstrated that the process can either strengthen or weaken pre-existing bars (also see \citealt{mi98}). \citet{mo17} and \citet{za17} showed that interactions with small companions can result in delayed bar formation. \citet{be04} found that bars regenerated in stellar discs, but not in dissipative discs. Instead the gas was efficiently funneled to the central regions. 

Thus, modeling to date suggests that the formation of long-lived stellar bars can triggered in interactions, but not gaseous bars. What about gaseous bars in isolated, but bar-unstable discs? Several published high resolution simulations in the literature partially address this question. For example, the models of \citet{ma04} produce small, eccentric, long-lived, gas bars, contained within larger stellar bars. The images of \citet{fi15} and \citet{sp17} suggest similar results, i.e., very small and weak gas bars, though even these don't appear as long-lived as in \citet{ma04}. \citet{fi15} report that gas is emptied rapidly in a dead zone between co-rotation and inner Lindblad resonances, making it hard to feed nuclear activity at later times. 

These high resolution results accord with several of the findings above for bars made of nested loop orbits. Specifically, that nested, non-intersecting orbits can exist in asymmetric external potentials, but that the most eccentric resonant orbits are relatively small, and that the strength of the asymmetric part of the external potential is relatively large. The decay of these model gas bars also agrees with the result that self-gravitating bars are unlikely to be stable and long-lived.

The model of \citet{re15} shows a younger bar than in most previous published simulations. The gaseous part of this bar consists of a thin elliptical annulus, with spiral-like waves on the inside and outside. The inner spirals meet the annular bar near the minor access of the latter. The morphology of this model gas bar suggests that it might consist of a group of orbits like those in Fig. 9, with crossings of between orbits outside a very narrow range of parameters resulting in spiral waves. 

To return to the case of interaction induced bars, the orbital results of the previous sections provide a basis for the following picture of tidal bar evolution. First, in a prolonged prograde encounter, the large scale coherence of the perturbing potential may excite nested eccentric resonant orbits with a common pattern speed, like those in Fig. 8. As explained in Sec. 5 the detailed structure of these orbits will depend on both the shape of the symmetric potential in the disc and the perturbing potential. In the gas, dissipation in nearby, non-resonant orbits may synchronize with the resonant orbit. The dissipation resulting from orbits much different than the resonant ones will drive radial flows, e.g., inside and outside an annular bar.  

When the companion galaxy leaves, or merges without substantially disrupting the bar, the external asymmetric force disappears, but the stars on resonant orbits may maintain a kinematic bar for some time. On the other hand, if the magnitude of the external potential was substantial, then its disappearance will perturb the orbits. If the bar has significant self-gravity, then it will have more crossing orbits, and dissipation in any remaining gas will aid its dissolution.  

\section {Summary and conclusions}

One major theme of this paper centres on the questions of when, or in what potentials, simple, closed loop orbits exist over a range of radii, as they evidently can in the well-studied Ferrer's potentials. When this is the case there exists a dense, nested ensemble of such orbits that could themselves serve as a simple model of the bar. Moreover, these loop orbits generally will be parents of more complex orbit families, as functions of the orbit parameters. A second theme concerns the usefulness of scale-free, power-law potentials, viewed as elementary forms from which more complex potentials could be constructed. The orbit structure of these simple potentials is less complex than many of those used in numerical simulations. In the above, we used p-ellipse functions to find and approximate the simplest orbits in these potentials, in a perturbation limit.

The p-ellipse approximation was originally found to provide very good fits to precessing, eccentric orbits in symmetric power-law potentials, as these fits do not drift with time \citep{st06}, since the approximation tracks the orbital precession. It is reasonable to expect that the approximation might also be useful for fitting resonant loop orbits (i.e., closed orbits in the pattern frame) in potentials with a modest asymmetric component, like a bar potential. Since in such orbits the ratio of the precession and pattern frequencies are rational numbers, it seems that any resonant orbit should have a modified precession rate that could be captured by a p-ellipse approximation.  

This conjecture was shown to be correct in Sec. 4, where closed orbits in the pattern frame were found to be well approximated by p-ellipses, see Figs. 1, 2, and 5. Good fits were found in potentials with both shallow and steeply rising rotation curves. Differences and similarities between the orbits in these two cases were discussed in Sec. 4.3. One of the most interesting results is that p-ellipse, loop (m = 2) orbits are not predicted to exist in (power-law)  falling rotation curve potentials by the perturbation analysis. However, loop orbits are found numerically in such potentials. The second-order p-ellipse solutions of Sec. 3 seem to fail in these potentials. This phenomenon will be explored in a later paper.

Even in the case of rising rotation curve potentials, the exploratory calculations of Sec. 4 show complexities.  Fig. 1 shows the closed, resonant orbit in a sample case. Fig. 2 shows that even a small deviation from the initial conditions of Fig. 1 yields a smaller, unclosed, and more eccentric orbit. Although it cannot capture the small librations of this orbit, the p-ellipse approximation is still a good fit to the mean orbit. Figs. 3 and 4 show that further small variations in the initial conditions yield progressively larger, and rounder orbits. These orbits are characterized by clear subharmonic frequencies. The p-ellipse approximation can only capture the innermost loops of these orbits. Given the proximity to the resonant orbit, the subharmonics may be the result of 'beating' between the precession and pattern frequencies. These subharmonics may also relate to the $\beta$ frequency in the epicyclic perturbation analysis of \citet[Sec. 3.3]{bt08}. The initial conditions around the simple closed loop are dense with resonances between the subharmonic and primary orbital frequency, which produce the closed multi-loop orbits like those in Figs. 3 and 4. Multi-frequency p-ellipse approximations to the closed, multi-loop orbits, using the equations of Sec. 3.2, will be investigated in a future paper. Although the resonant, multi-loop orbits cannot be modeled by simple, p-ellipses, solutions to the p-ellipse constraint equations guide us to good estimates of their initial conditions by simply varying the parameter $f_o$. 

Resonant orbits from a small region of parameter space, like those in Figs. 1-4, can be combined to produce a model bar in the given potential.  This provides a different technique to the more traditional one of constructing orbits via perturbation ellipses around Lagrange points in a given potential (see \citealt[Sec. 3.3]{bt08}). 

Figs. 4 and 8 show a surprising feature of closed, multi-loop orbits. Half of their innermost loops can be well fit by a p-ellipse with nearly the same initial conditions, and these segments could support the bar. However, their outermost loops are generally much more circular, and in isolation would look like segments of a disc orbit unrelated to the bar.

In a given power-law potential (specified by the values of $\delta$ and $\delta_b$) each approximate, resonant orbit requires a specific value of the asymmetric amplitude $e_b$ to satisfy the equation of motion constraints. For an ensemble of such orbits, making up a model bar, a specific radial variation of $e_b$ is required. It is unlikely that the needed pattern of $e_b$ would obtain for arbitrary initial disc structures. However, gas clouds will prefer more or less concentric loop orbits. Dissipative processes might drive bar parameters (e.g., a combination of external and internal contributions to the asymmetric potential) to values that support the required variation. 

Note that the constraint equations are such that the required values of $e_b$ do not depend on the pattern speed. The zeroth-order orbital frequency parameter, $f_o$, does. An alternative, would be radial variations of the potential profile index ($\delta$, or $\delta_b$). Such variations must be small, or the perturbation equations have to be modified. An example was discussed in Sec. 5, and illustrated in Fig. 8. If, as in this example, the rotation curve becomes flat or slightly declining, subharmonic frequencies become dominant, and we get large, multi-loop orbits.  

Going beyond external asymmetric potentials, in the construction of a model self-gravitating bar the Poisson eq. can be viewed as a prescription for the density distribution of concentric loop orbits making up the bar (see Sec. 6.1). It also imposes strong additional constraints in a perturbation approximation. 

In fact, orbital solutions to the perturbation equations with the additional constraints from the Poisson equation given in Sec. 6.1 seem to be very rare. Evidently two-dimensional  bars must be based on more complex orbits, or perhaps most bars have a three dimensional structure.

Altogether, these results suggest a number of interesting general conclusions. 1) Perhaps the most important of these is that it is hard to make model bars from single loop orbits alone, and thus, bars and oval distortions made primarily from gas are unlikely to form in galaxy discs. This conclusion is not too surprising since bars are observed to be primarily stellar, and an extensive literature of models shows that they generally have a wide range of orbits, including chaotic ones (e.g., \citealt{co02, we15a, we15b} and references therein). On the other hand, the analytic and numerical explorations above provide some insights as to why this is so. These insights include the fact that resonant, loop orbits (approximated by p-ellipses) are only found in limited regions of parameter space, at least in the perturbation approximations of Sec. 3. 

2) In non-self-gravitating cases, these solutions tend to have large values of the asymmetric amplitude $e_b$, suggesting the external potential has a strong asymmetric part. Technically, this violates one of the perturbation approximations, which assumed that $e_b \simeq e$, but the good fits between numerical and analytic orbits at low to moderate values of $e$ suggests the consequences are not serious. 

3) The bar orbits illustrated above are wide, even at quite high values of the eccentricity parameter $e$. This suggests that more complex orbits (e.g., like those in described in \citealt{wi17}) are needed to support narrow bars.  

4) Loop orbit solutions in two-dimensional self-gravitating bars are at best very rare. It is not surprising that significant self-gravity would lead to more complex orbits. This may be a factor in understanding why bars are effective at quenching gas-rich discs (e.g., \citealt{kh17}). 

Beyond this initial exploration, there are many directions to pursue using p-ellipse approximation tool for the study of orbits in galaxies aysymmetric potentials. For example, p-ellipses approximations with multiple frequencies likely converge much more quickly than conventional Taylor expansions in $cos(\phi)$. Specifically, the case of librating p-ellipse orbits with an additional subharmonic frequency will be described in a later paper.

\section*{Acknowledgments}

I am very grateful for the insights gained from a correspondence over the last decade on orbits in galaxies with the late Donald Lynden-Bell. I acknowledge use of NASA's Astrophysics Data System.

\bibliographystyle{mn2e}

\bsp
\label{lastpage}
\end{document}